# Terahertz and infrared links in controlled non-Kolmogorov turbulence

**Hanchen Liu,[1] He Ni,[1] Xiaoxiang Li,[1] Jiabiao Zhao,[1] Yuheng Song,[1] Lingfeng Tian,[1] Yi Luo,[1] and Jianjun Ma[1,2*]**

[1]*School of Integrated Circuits and Electronics, Beijing Institute of Technology, Beijing 100081, China*
[2]*State Key Laboratory of Environment Characteristics and Effects for Near-space, Beijing 100081, China*

**Abstract:** Hybrid terahertz/infrared (THz/IR) links are usually modeled with Kolmogorov turbulence, but whether the turbulence spectral type matters equally for both bands has not been investigated. We transmit co-propagating 625 GHz and 1550 nm links through laboratory-controlled turbulence while a Mach-Zehnder interferometer and thermistors independently measure its spectral exponent and structure parameter on the shared path. Inlet-air temperature, not airflow speed, drives the exponent from Kolmogorov-like to non-Kolmogorov. At comparable structure parameters, this change reverses which model better reproduces the IR attenuation at the nominal inner scale, whereas the THz attenuation is insensitive to it. With type-matched models, predicted and measured bit-error-rate trends agree within 0.14 (IR) and 0.06 (THz) decades. IR predictions therefore require the spectral shape and inner scale, not turbulence strength alone.
**Keywords:** Atmospheric turbulence; non-Kolmogorov spectrum; terahertz communications; infrared communications; hybrid terahertz/infrared links

## 1. Introduction

Sixth-generation (6G) backhaul and fronthaul require wireless links with near-fiber capacity that remain available in adverse weather [1,2]. Hybrid terahertz/infrared (THz/IR) links address this need because the two bands are complementary. Free-space optical (FSO) links at 1550 nm offer Tb/s-class capacity over license-free, highly directional beams [3,4] but are strongly attenuated by fog and dust and suffer turbulence-induced fading [4-6]. THz carriers offer tens of gigahertz of bandwidth, are far less affected by fog and dust, and are less susceptible to scintillation [5-7], although rain attenuates both bands comparably [8] and spreading loss and molecular absorption limit the THz range [2,9]. Seamless switching between 1550 nm and 140 GHz links has been demonstrated in fog [10], and the wider THz beam could assist the acquisition, pointing, and tracking of narrow IR beams between moving nodes [11,12]. Exploiting this complementarity requires experimentally validated models that connect the turbulence experienced by both bands to link attenuation and bit error rate (BER).

Refractive-index fluctuations distort the wavefront and produce irradiance fluctuations that cause fading and raise the BER [13]. Because the THz wavelength is two to three orders of magnitude longer than the IR wavelength, a given optical-path fluctuation perturbs the THz phase far less; a co-propagating laboratory comparison accordingly found IR attenuation approaching 1 dB but THz attenuation of only tens of millidecibels [14]. Most link models assume the Kolmogorov spectrum with its 11/3 inertial-range exponent [13]. However, non-Kolmogorov scaling has been reported in the stable boundary layer [15], over near-ground optical paths [16] and in interferometric phase and wavefront measurements [17,18], in controlled thermal convection [19], and in the free troposphere and stratosphere [20-22]. These findings motivated layered non-Kolmogorov propagation models [23] and non-Kolmogorov analyses of satellite uplinks [24], and high-altitude-platform links likewise depend on high-altitude turbulence statistics [25]. Generalized non-Kolmogorov spectra, finite-aperture Gaussian-beam scintillation models [26-29], and channel models built on them [30,31] describe these departures and predict capacity and error rates, but they prescribe the spectral exponent, turbulence strength, and irradiance statistics. Whether their predictions hold when independently measured spectral and strength parameters are used as inputs remains largely untested, and whether the spectral type matters equally for THz and IR carriers sharing one turbulent path has, to our knowledge, not been examined experimentally.

In this article, we address both questions with co-propagating 625 GHz and 1550 nm links in laboratory-controlled turbulence. A Mach-Zehnder interferometer (MZI) and thermistors on the shared path independently measure the effective spectral exponent $\alpha$ and the refractive-index structure parameter $C_n^2$, which serve as inputs to Kolmogorov (K) and generalized non-Kolmogorov (NK) finite-aperture propagation models. Comparing their predictions with synchronized attenuation and BER measurements tests the chain from turbulence spectrum through scintillation and attenuation to BER.

## 2. Experimental setup and turbulence characterization

### *2.1 Experimental setup*

The experimental platform in Fig. 1 is built around the 625 GHz/1550 nm dual-band testbed. Figure 1(a) shows the co-propagating transmission path and the power- and BER-measurement branches for both bands, and Fig. 1(b) shows the turbulence chamber, temperature sensors, and MZI. In the turbulence chamber, air enters through two top ports and exits through a continuous slot along the lower front side. Four conditions combine warm (55 °C) or hot (70 °C) inlet air with low (28.6 m/s) or high (41.6 m/s) nominal airflow speed. Each condition is analyzed from a single record with a 0–10 s no-flow baseline and airflow applied during 10–20 s. The MZI and IR-monitoring signals are sampled at 125 kHz. Received powers and BERs are recorded at a minimum interval of 500 ms. All records are aligned to a common time axis and analyzed in one-second blocks. The THz power-meter resolution is 0.005 dB. Because the communication channels and the phase-probe beam traverse the same turbulent region, the turbulence parameters and link performance are measured simultaneously in the same turbulence flow.

The THz transmitter employs a 625 GHz source based on a Schottky diode frequency multiplier chain with a continuous wave output power of approximately 1 mW (Fig. 1(e)) [32]. The radiation is emitted through a horn antenna with a 2.4-mm aperture and collimated by a lens with a 32 mm focal length into a beam approximately 15 mm in diameter. After an approximately 3 m free space path, it is coupled into the THz receiver through an identical lens and horn. A 2.5 Gb/s data pattern is transmitted with duobinary modulation to reduce the signal bandwidth. The infrared (IR) transmitter comprises a 1550 nm distributed feedback laser, an erbium doped fiber amplifier, an optical filter with an approximately 1 nm bandwidth, a variable optical attenuator, a polarization controller, and a Mach-Zehnder modulator, and it uses the same data pattern with non-return-to-zero on-off keying (NRZ-OOK). The IR beam is expanded to approximately 15 mm and combined with the THz beam by an approximately 2-μm-thick pellicle beam splitter, and both beams propagate through the turbulence chamber along a shared path. An adjustable iris sets the received power so that the BER without imposed turbulence is on the order of $10^{-7}$ [14].

At the receiver, a beam separator directs the IR signal into the IR branch, where a tap sends part of the optical power to the free-space detector PD_FS (effective detection area 19.6 $mm^2$) and the remaining power is coupled through a fiber collimator into a 1×2 multimode-fiber (MMF) splitter. One splitter output feeds the IR data receiver, which measures the received power and performs data detection, while a bit-error-rate tester records the BER. The other output is divided by a second 1×2 MMF splitter between PD_SSMF and PD_MMF, which monitor the power through a standard single-mode fiber (approximately 9 μm core) and a multimode fiber (approximately 50 μm core), respectively, for comparison at different collection scales. The THz signal is coupled through the receiving lens and horn into the THz receiver, whose received power and BER are recorded synchronously. The effective IR receiver scale used in the propagation calculations is associated with the fiber-coupled data branch.

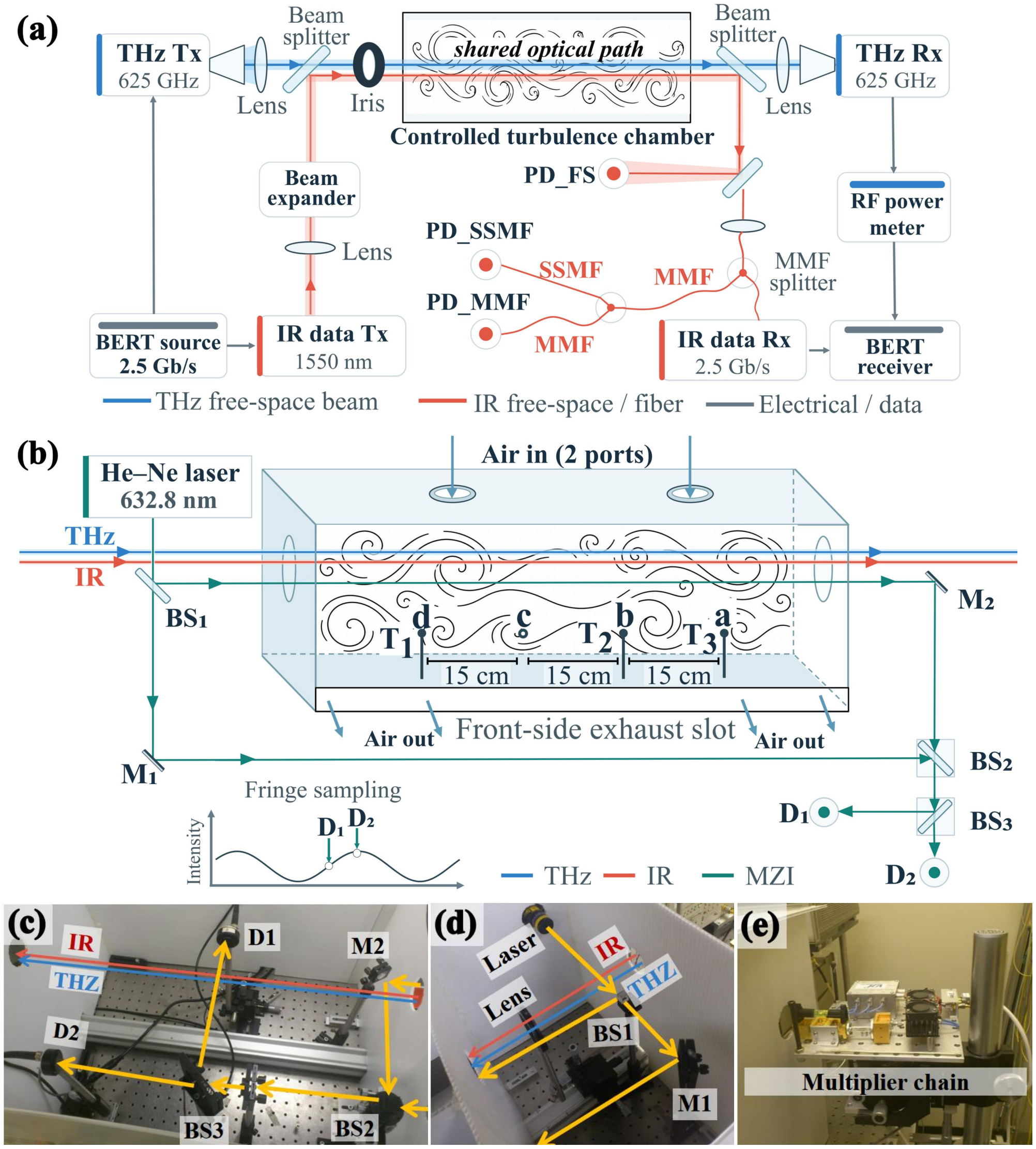

Fig. 1. Experimental platform for the co-propagating 625 GHz THz and 1550 nm IR links. (a) Dual-band communication setup; (b) controlled-turbulence chamber and MZI configuration; (c,d) photographs of the chamber-exit and chamber-entry MZI optics, respectively; (e) 625 GHz THz transmitter.

### *2.2 Turbulence characterization*

The MZI phase of the 632.8 nm He–Ne probe is recovered from the $D_1$ and $D_2$ signals with a nonorthogonal I/Q correction [33,34], which removes the channel offsets, the gain mismatch, and the departure of the channel phase separation from 90° (Supplement 1, Section S1). Its temporal power spectral density (PSD) is estimated with Welch's method [35,36]. Under the Taylor frozen-flow approximation [37-40], with transverse statistical homogeneity and local isotropy, a temporal phase PSD $S_\varphi(f) \propto f^{\beta}$ corresponds to a refractive-index spectrum $\Phi_n(\kappa) \propto \kappa^{-\alpha}$ with $\alpha = \beta + 1$ [41] (Supplement 1, Section S2). We fit $\beta$ over a common wavenumber window $\kappa$ = 756-2459 rad/m, mapped from frequency with the nominal airflow speeds. The lower bound excludes recurring fixed-frequency interference, and the upper bound excludes the dissipation-range roll-off [42,43] (Supplement 1, Table S1). Within this window, the spectral shapes measured at the two airflow speeds collapse, as expected for a velocity-independent spatial spectrum (Supplement 1, Fig. S3), and Welch-segment bootstrap resampling gives the 95% confidence interval (CI) of each stage-level $\alpha$.

Linear fits to the smoothed mean of the three temperature records give warming rates of 1.26-3.42, 0.71-1.63, and 0.15-0.47 °C/s over 10-12, 12-15, and 15-20 s (ranges span the four conditions), defining the initial-establishment, development, and quasi-steady stages. The cross-speed spectral collapse is closest in the quasi-steady stage, which supports this segmentation.

**Table 1. Stage-level spectral exponent $\alpha$ with its 95% bootstrap CI and $C_n^2$ (in $10^{-10}$ $m^{-2/3}$) within the common $\kappa$ window.**

| Condition | 10-12 s | | 12-15 s | | 15-20 s | |
|---|---|---|---|---|---|---|
| | $\alpha$ [95% CI] | $C_n^2$ | $\alpha$ [95% CI] | $C_n^2$ | $\alpha$ [95% CI] | $C_n^2$ |
| Warm/low | 3.95 [3.77, 4.18] | 3.78 | 3.60 [3.55, 3.65] | 5.67 | 3.56 [3.49, 3.59] | 6.82 |
| Warm/high | 3.92 [3.85, 4.04] | 7.09 | 3.70 [3.63, 3.74] | 10.8 | 3.62 [3.58, 3.65] | 11.6 |
| Hot/low | 3.47 [3.31, 3.91] | 6.02 | 3.35 [3.30, 3.43] | 10.3 | 3.34 [3.29, 3.38] | 13.2 |
| Hot/high | 3.50 [3.47, 3.62] | 15.5 | 3.48 [3.42, 3.52] | 23.8 | 3.39 [3.35, 3.42] | 24.4 |

The CIs in Table 1 are widest in the nonstationary initial-establishment stage and narrow markedly thereafter. In the developed stages (12-20 s), $\alpha$ = 3.56-3.70 under the warm and 3.34-3.48 under the hot conditions. At a fixed airflow speed, the warm–hot differences (0.22-0.25) exceed the estimation uncertainty, because the CIs do not overlap; at a fixed temperature, the speed-induced differences (0.05-0.13) are smaller and their CIs overlap. The inlet-air temperature therefore controls the departure from Kolmogorov scaling: $|\alpha - 11/3|$ averages 0.07 under the warm but 0.28 under the hot conditions. Because even the quasi-steady warm CIs lie slightly below 11/3, and the inner scale, outer scale, and transverse statistics are not measured independently, this classification is relative: we describe the warm turbulence as Kolmogorov-like and the hot turbulence as non-Kolmogorov.

Turbulence strength is characterized by the K-equivalent structure constant $C_n^2$ and the generalized NK spectral-strength parameter $\tilde{C}_n^2$ [13,26,44]. Because the refractive-index structure quantity $D_n(r_T)$ is measured only at the thermistor-pair separation $r_T$ = 0.15 m (Supplement 1, Section S3), we follow the fixed-scale structure-function matching of Ref. [44] and set $L_{\text{ref}} = r_T$ in

$$C_n^2 = D_n(r_T) r_T^{-2/3}, \qquad \tilde{C}_n^2 = D_n(L_{\text{ref}}) L_{\text{ref}}^{3-\alpha} = C_n^2 L_{\text{ref}}^{11/3-\alpha} \tag{1}$$

so that the K and NK structure functions have the same amplitude at the measured scale. $\tilde{C}_n^2$ ($m^{3-\alpha}$) is used in the NK calculations, whereas $C_n^2$ ($m^{-2/3}$) represents turbulence strength in all tables and figures. The same $C_n^2$ is applied to both bands, assuming temperature-dominated fluctuations and neglecting humidity fluctuations, to which the THz refractive index is relatively more sensitive. Repeating the time alignment, baseline correction, and $D_n$ estimation for thermistor response times $\tau$ = 2.5-3.5 s around the nominal 3 s gives the $C_n^2$ uncertainty used for the prediction error bars in Section 4. $C_n^2$ increases with both the inlet-air temperature and the airflow speed (Table 1). The one-second $\alpha$ and $C_n^2$ values are listed in Supplement 1, Tables S2-S5.

## 3. Propagation Modeling

### *3.1. Scintillation and attenuation*

The NK scintillation index is calculated numerically with the finite-aperture Gaussian-beam model [28] using the generalized modified spectrum $\Phi_n(\kappa, \alpha)$ of Ref. [26], which includes diffraction, aperture averaging, and finite inner- and outer-scale cutoffs:

$$
\sigma_{I,\mathrm{NK}}^{2}\left(D_{\mathrm{G}}\right)=8\pi^{2}k^{2}L\int_{0}^{1}\int_{0}^{\infty}\kappa\,\Phi_{n}\left(\kappa,\alpha\right)\exp\left\{-\frac{L\kappa^{2}}{k\left(\Omega_{\mathrm{G}}+\Lambda\right)}\left[\left(1-\bar{\Theta}\xi\right)^{2}+\Lambda\Omega_{\mathrm{G}}\xi^{2}\right]\right\}
$$
$$
\times\left\{1-\cos\left[\frac{L\kappa^{2}}{k}\left(\frac{\Omega_{\mathrm{G}}-\Lambda}{\Omega_{\mathrm{G}}+\Lambda}\right)\xi\left(1-\bar{\Theta}\xi\right)\right]\right\}\mathrm{d}\kappa\,\mathrm{d}\xi \qquad (2)
$$

Here $\kappa$ and $k$ are the spatial and optical wavenumbers, $L$ is the path length, $D_G$ is the receiver-aperture diameter, and $\xi = 1-z/L$ is the normalized path coordinate. For an input beam of effective radius $W_0$ and wavefront radius of curvature $R_0$, $\Theta_0 = 1-L/R_0$ and $\Lambda_0 = 2L/(kW_0^2)$. At the receiver, $\Theta = \Theta_0/(\Theta_0^2+\Lambda_0^2)$, $\Lambda = \Lambda_0/(\Theta_0^2+\Lambda_0^2)$, $\bar{\Theta} = 1-\Theta$, and $\Omega_G = 16L/(kD_G^2)$. The K baseline is the finite-aperture plane-wave scintillation index for weak-to-moderate Kolmogorov turbulence [13,45],

$$
\begin{aligned}
\sigma_{I,\mathrm{K}}^{2}(D) &= \exp\left[\sigma_{\ln X}^{2}(D)+\sigma_{\ln Y}^{2}(D)\right]-1,\\
\sigma_{\ln X}^{2}(D) &= \frac{0.49\sigma_{\mathrm{R}}^{2}}{\left(1+0.65d^{2}+1.11\sigma_{\mathrm{R}}^{12/5}\right)^{7/6}},\\
\sigma_{\ln Y}^{2}(D) &= \frac{0.51\sigma_{\mathrm{R}}^{2}\left(1+0.69\sigma_{\mathrm{R}}^{12/5}\right)^{-5/6}}{1+0.90d^{2}+0.62d^{2}\sigma_{\mathrm{R}}^{12/5}}.
\end{aligned} \qquad (3)
$$

where $\sigma_{\ln X}^2$ and $\sigma_{\ln Y}^2$ are the large- and small-scale log-irradiance variances, $\sigma_R^2 = 1.23C_n^2k^{7/6}L^{11/6}$ is the Rytov variance, and $d = (kD^2/4L)^{1/2}$. Because the two models differ in spectral normalization, wave model, inner-scale treatment, and aperture kernel, setting $\alpha = 11/3$ does not make Eqs. (2) and (3) equivalent. An NK control with $\alpha$ fixed at 11/3 therefore separates the effect of the exponent from that of the model framework (Section 4). Following Ref. [14], the scintillation index is mapped to an equivalent attenuation,

$$
A_{\mathrm{turb}}=-10\log_{10}[1-\sqrt{\sigma_I^{2}(D)}] \qquad (4)
$$

an empirical metric that expresses a one-standard-deviation fade in decibels. Because weak scintillation redistributes rather than removes power, Eq. (4) is a fade metric rather than a power loss and is applied identically to both models

Table 2 lists the model inputs. The beams are treated as collimated ($R_0 = \infty$) with $W_0 = 7.5$ mm, matching the ~15 mm beam diameters, and the effective turbulence length $L_{\mathrm{eff}} = 0.6$ m inferred for this chamber in Ref. [14] represents the cumulative effect of the nonuniform turbulence. The THz receiver diameter is the 2.4 mm horn aperture. The IR receiver is represented by a model-equivalent diameter $D_{IR} = 50$ μm, the spatial acceptance scale of the MMF-coupled data branch, which maps the fiber-coupled receiver onto the aperture-averaging variable of the NK model [28]. Fixed coupling losses cancel because the attenuation is referenced to the no-flow baseline. Because neither the inner scale $l_0$ nor the outer scale $L_0$ was measured, the calculations are repeated over $l_0 = 1.0$-$2.0$ mm and $L_0 = 0.09$-$20$ m, spanning the ranges reported in the literature [46-48].

Both bands satisfy $\sigma_R^2 \ll 1$ (Table 2) and thus lie in the weak-fluctuation regime of Eqs. (2) and (3), but they sample different parts of the turbulence spectrum. The IR beam propagates as a nearly plane wave whose Fresnel scale lies near the inner scale, so its scintillation is governed by eddies near the dissipation range. The THz beam diffracts appreciably within the chamber (Rayleigh range ≈ 0.37 m), and its Fresnel scale samples eddies an order of magnitude larger. For a common $C_n^2$, the plane-wave Rytov variances of the two bands differ by $(\lambda_{THz}/\lambda_{IR})^{7/6} \approx 805$. These scale differences underlie the band-dependent model behavior in Section 4.

Table 2. Model inputs and band-dependent propagation scales for $L = L_{eff}$; $\sigma_R^2$ is evaluated over $C_n^2$ = $3.78\times10^{-10}$–$2.44\times10^{-9}$ m$^{-2/3}$ (Table 1).

| Quantity | IR (1550 nm) | THz (625 GHz) |
|---|---|---|
| Effective turbulence length $L_{eff}$ | 0.6 m | |
| Inner scale $l_0$ (scan range) | 1.5 mm (1.0-2.0 mm) | |
| Outer scale $L_0$ (scan range) | 0.9 m (0.09–20 m) | |
| Reference scale $L_{ref} = r_T$ | 0.15 m | |
| Input-beam radius $W_0$; wavefront curvature $R_0$ | 7.5 mm; $\infty$ | |
| Receiver diameter $D$ | 50 µm (model-equivalent) | 2.4 mm |
| Rytov variance $\sigma_R^2$ | 0.0093–0.060 | $(1.16\text{-}7.48)\times10^{-5}$ |
| Fresnel scale $(L/k)^{1/2}$ | 0.385 mm | 6.77 mm |
| Fresnel wavenumber $(k/L)^{1/2}$ | $2.6\times10^3$ rad/m | 148 rad/m |
| Inner-scale parameter $Q_l = 10.89L/(kl_0^2)$ | 0.72 | 222 |
| Fresnel ratio $\Lambda_0 = 2L/(kW_0^2)$ | $5.3\times10^{-3}$ | 1.63 |

### *3.2. Irradiance statistics and BER*

The average BER is obtained by averaging the conditional BER over the normalized received irradiance $I$ (unit mean, variance $\sigma_I^2$) [13,49,50]:

$$\bar{P}_b\left(\bar{\gamma},\sigma_I^2\right)=\int_0^\infty P_b\left(\bar{\gamma},I\right)p_I(I)\,\mathrm{d}I \tag{5}$$

where $\bar{\gamma}$ is the mean electrical signal-to-noise ratio (SNR) without fading, determined from the received signal power without turbulence and the receiver noise power measured with the transmitter switched off, and $p_I(I)$ is the irradiance probability density. Because both links are in the weak-fluctuation regime, we use the lognormal density of first-order Rytov theory [13,49,51,52],

$$p_{\mathrm{LN}}(I)=\frac{1}{I\sqrt{2\pi\sigma_X^2}}\exp\left[-\frac{(\ln I-\mu_X)^2}{2\sigma_X^2}\right] \tag{6}$$

with $\sigma_X^2 = \ln(1 + \sigma_I^2)$ and $\mu_X = -\sigma_X^2/2$, and test it against a single-parameter Gamma density with the same $\sigma_I^2$ [24,53] (Section 4.3). The conditional BER kernels of the OOK intensity-modulation/direct-detection (IM-DD) IR link [49] and of the duobinary THz link are

$$P_{b,\mathrm{IR}}\left(\bar{\gamma},I\right)=Q\left(\sqrt{\bar{\gamma}}\,I\right),\qquad P_{b,\mathrm{THz}}\left(\bar{\gamma},I\right)=C_{\mathrm{DB}}Q\left(\sqrt{\eta_{\mathrm{DB}}\bar{\gamma}I}\right) \tag{7}$$

where $Q(x)=erfc\left(x/\sqrt{2}\right)/2$, and $C_{\mathrm{DB}}$ and $\eta_{\mathrm{DB}}$ are an amplitude normalization and an equivalent SNR-scaling coefficient of the duobinary kernel, both set to unity to compare the relative BER trends produced by different irradiance statistics. For THz, we approximate the effective SNR as proportional to irradiance under weak fading (the calculated BER differs by less than 0.0031 decades from that obtained using square-law SNR scaling, with all other inputs fixed).

## 4. Analysis and results

### *4.1. Infrared Attenuation*

The measured attenuation is the decrease in received power relative to the 0-10 s no-flow baseline in each one-second block. The predictions use the one-second $C_n^2$ and, for the NK model, the one-second $\alpha$ and $\tilde{C}_n^2$. Initial-establishment blocks whose $\alpha$ lies outside the NK domain $3 < \alpha < 4$ have no NK prediction. Model errors are quantified by the mean absolute error (MAE) over the blocks for which both predictions exist.

Under the warm, Kolmogorov-like conditions (Fig. 2(a,b)), the K model is clearly better (K/NK MAEs of 0.0315/0.2135 and 0.0841/0.2539 dB). The NK predictions lie systematically below the measurements but track their block-to-block variation (fitted slopes 0.97-1.11), whereas the K predictions match the mean level but underestimate the variation (slopes 0.48-0.68). The K advantage thus lies in the absolute level, which also depends on the thermistor-derived $C_n^2$ and on $L_{\mathrm{eff}}$.

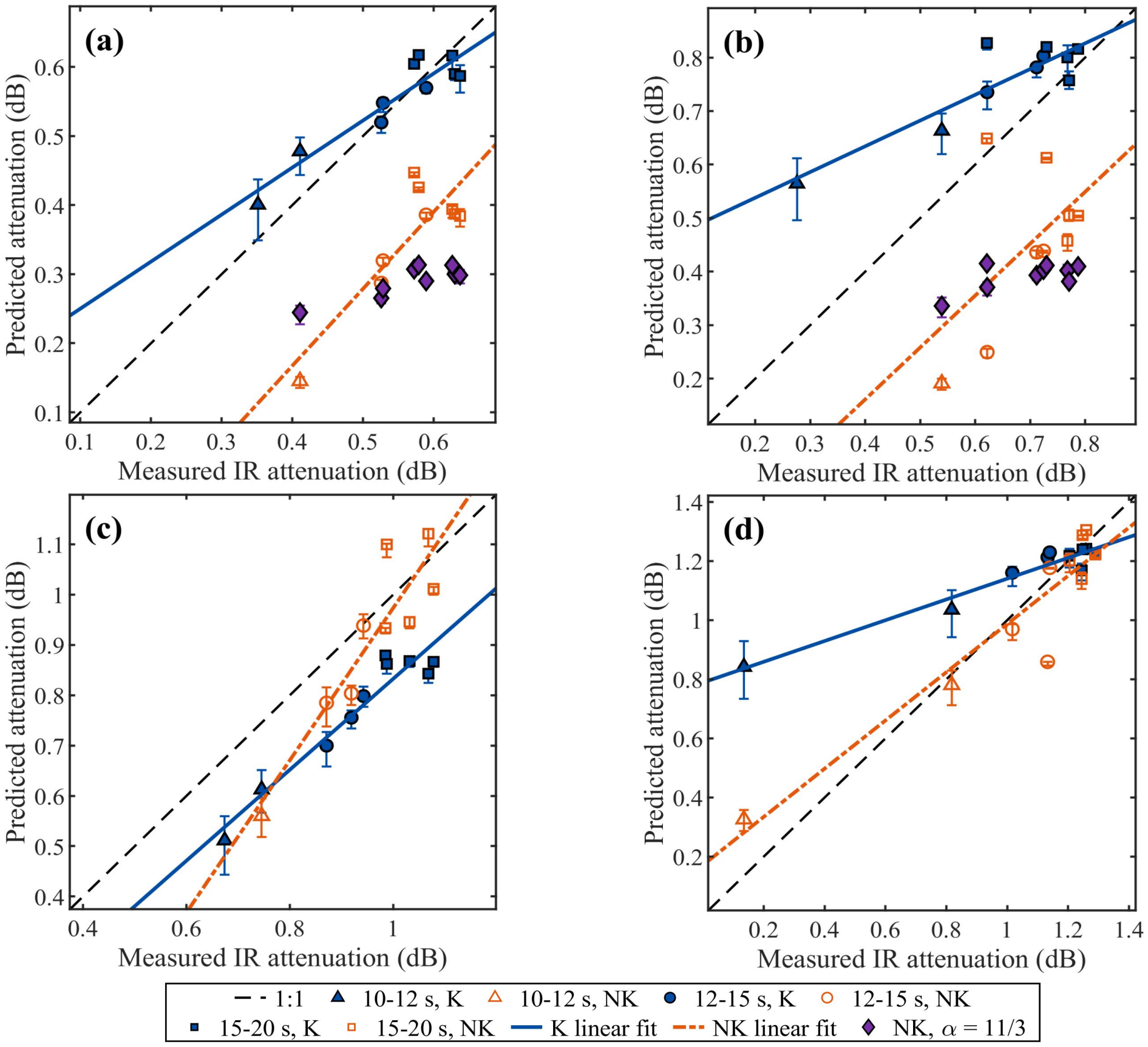

Fig. 2. One-second measured IR attenuation versus the K and NK predictions over the three stages for (a) warm/low, (b) warm/high, (c) hot/low, and (d) hot/high conditions. Purple diamonds denote the NK control with $\alpha = 11/3$ (warm conditions). Error bars denote the prediction range for $C_n^2$ recalculated with $\tau = 2.5$-$3.5$ s.

The NK control locates the origin of the NK offset: with $\alpha = 11/3$, the NK MAEs increase to 0.2763 and 0.3056 dB, so the offset stems not from the measured exponent but from the inner-scale treatment. For $\Lambda_0 \approx 5\times10^{-3}$ the Gaussian-beam and plane-wave kernels nearly coincide, but the NK spectrum includes the inner-scale cutoff and bump [26], which Eq. (3) neglects. At $Q_l \approx 0.72$, this cutoff removes much of the spectral power that drives IR scintillation: weak-fluctuation plane-wave theory with the modified spectrum gives $\sigma_I^2 \approx 0.26\sigma_R^2$ instead of $\sigma_R^2$ for $l_0 = 0$ [13], roughly halving the predicted attenuation. The lower K-model error is thus consistent with, but does not by itself prove, the Kolmogorov-like classification.

Under the hot conditions, $\alpha$ remains below 11/3 in every stage (Table 1), and the NK model gives the lower error at the nominal $l_0$. For hot/low (Fig. 2(c)), the NK model gives the lower MAE in both the development and the quasi-steady stage (K/NK MAEs of 0.1596/0.0684 and 0.1662/0.0737 dB, respectively). For hot/high (Fig. 2(d)), the 10–20 s K/NK MAEs are 0.1417/0.0848 dB, a 40.1% reduction that arises mostly from the two 10–12 s blocks. From 12 to 20 s, the two predictions nearly coincide. The error bars do not change these type-dependent rankings.

Within the NK model, the hot-condition advantage reflects the separation between the $C_n^2$ reference scale and the eddies that drive IR scintillation. With the spectra matched at $L_{\mathrm{ref}}$ = 0.15 m [Eq. (1)], their ratio at wavenumber $\kappa$ is approximately $[A(\alpha)/A(11/3)](\kappa L_{\mathrm{ref}})^{11/3-\alpha}$, where $A(\alpha) = \Gamma(\alpha\text{-}1)\cos(\pi\alpha/2)/(4\pi^2)$ [26] and inner- and outer-scale factors are neglected. At the IR Fresnel wavenumber ($2.6\times10^3$ rad/m), near both the inner-scale cutoff $\kappa_l$ and the upper edge of the $\alpha$ window, this ratio is 1.2-1.6 for the quasi-steady warm exponents but 2.9-3.3 for the hot ones. A lower $\alpha$ thus transfers spectral power from the reference scale to the millimeter scales that govern IR scintillation and, through the cutoff relation of Ref. [26], raises $\kappa_l$ from $2.3\times10^3$ to $2.9\text{-}3.0\times10^3$ rad/m. Together, these effects offset the inner-scale suppression and raise the NK predictions toward the measurements, consistent with the scintillation enhancement predicted for $\alpha < 11/3$ [54]. Because the IR irradiance variance depends on $l_0$ at fixed $C_n^2$ when $Q_l \sim 1$ [13,55], the same balance makes the hot-condition ranking sensitive to $l_0$: the K model remains better under the warm conditions over the whole $l_0$ scan, whereas the NK model is preferred under the hot conditions only for $l_0 \approx$ 1.26-1.71 mm (hot/low) and 1.36-1.87 mm (hot/high), about half of the scan (Supplement 1, Section S4, Table S6, and Fig. S4). Outer-scale variations change the IR predictions by <0.1% (Supplement 1, Fig. S5). Notably, the quasi-steady warm/high and hot/low conditions have similar $C_n^2$ ($1.16\times10^{-9}$ and $1.32\times10^{-9}$ $\mathrm{m}^{-2/3}$) but opposite rankings (Fig. 2(b,c)): at the nominal $l_0$, the spectral type rather than the turbulence strength determines which model applies.

### *4.2 THz Attenuation*

Across all conditions and stages, the measured THz attenuation is only 0.005-0.030 dB, one to six display-resolution steps of the power meter, versus up to ~1.3 dB for IR (Figs. 2 and 3). This contrast is of the order expected from the Rytov-variance ratio of ≈805, i.e., a ≈28-fold smaller plane-wave scintillation standard deviation, and agrees with earlier co-propagation measurements [14]. The NK model gives the smaller THz error under all four conditions, including the warm ones, also within the $C_n^2$ uncertainty. The warm-condition NK control resolves this apparent conflict with the IR ranking: the NK predictions with $\alpha = 11/3$ and with the measured $\alpha$ nearly coincide but remain well below the K predictions (Fig. 3(a,b)). The THz advantage of the NK model therefore stems from its Gaussian-beam framework rather than from the spectral exponent, although the K-NK separation spans only one to three resolution steps: with $\Lambda_0 \approx 1.6$, weak-fluctuation theory gives an on-axis scintillation index of only $\approx 0.25\sigma_R^2$, versus $\sigma_R^2$ for the plane-wave K baseline [13], consistent with K predictions that generally exceed the measurements. In contrast, $\alpha$ has little influence at 625 GHz: the THz Fresnel scale far exceeds $l_0$ ($Q_l \approx 222$), its Fresnel wavenumber (148 rad/m) lies below the $\alpha$ window, and the NK/K spectral ratio there is only 1.07-1.29. Outer-scale variations change the THz predictions by up to 6.2%, which remains below the 0.0025 dB half-step quantization interval in absolute terms (Supplement 1, Fig. S5). The THz ranking thus reflects beam diffraction and does not contradict the MZI-based classification.

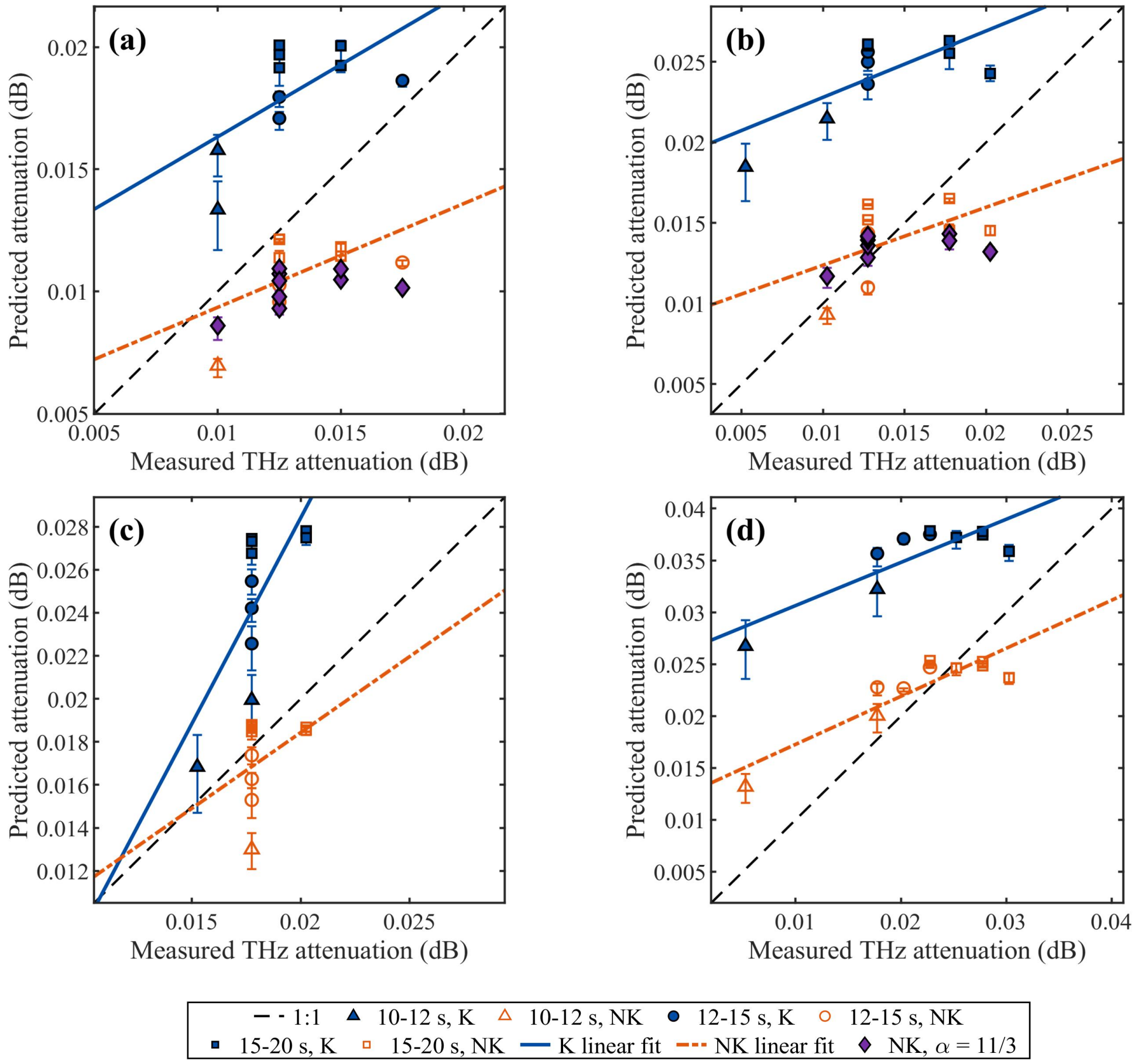

Fig. 3. One-second measured THz attenuation versus the K and NK predictions over the three stages for (a) warm/low, (b) warm/high, (c) hot/low, and (d) hot/high conditions. Purple diamonds denote the NK control with $\alpha = 11/3$ (warm conditions). Error bars denote the prediction range for $C_n^2$ recalculated with $\tau = 2.5$-$3.5$ s.

### *4.3. End-to-end BER prediction*

For IR, lognormal averaging [Eqs. (5)-(7)] reproduces the measured BER–SNR dependence over the measured mean-SNR interval (12.82-14.25 dB) with a root-mean-square error (RMSE) of 0.138 decades, lower than Gamma averaging with the same kernel and SNR calibration (Supplement 1, Section S5 and Fig. S6). For the same $\sigma_I^2$, the Gamma density decays only as a power law toward $I \rightarrow 0$ and thus assigns more probability to deep fades, overestimating the BER; fade predictions are known to be sensitive to this tail [56], and the lognormal agreement is consistent with first-order Rytov theory and measured weak-turbulence statistics [51,52]. For THz ($\sigma_R^2 < 10^{-4}$), both densities collapse toward $I = 1$ and give nearly identical BERs, so the THz data cannot discriminate between them; the lognormal density is used for both links.

Combining the propagation models selected in Sections 4.1 and 4.2 with lognormal statistics completes the chain from ($\alpha$, $C_n^2$) through $\sigma_I^2$, attenuation, and SNR to BER (Fig. 4). NK-modeled blocks outside $3 < \alpha < 4$ are excluded from both the calculated and the measured sets. For IR, with the K model for the warm and the NK model for the hot conditions, the calculated and measured BER trends both rise by about two orders of magnitude and largely overlap, differing by at most 0.141 decades (median ≈0.05 decades) within their common $C_n^2$ interval. Because each condition's model was selected from the attenuation comparison on the same records, this agreement is a consistency check of the full chain rather than an independent validation. It is nevertheless informative: near the operating point the IR BER curve is steep (Supplement 1, Fig. S6(a)), so K-NK attenuation-MAE differences of up to ≈0.18 dB (Section 4.1) would likely produce BER errors exceeding this agreement. For THz, the NK model is used throughout. The calculated and measured trends differ by 0.03-0.06 decades (maximum 0.059; median ≈0.056), with the measured trend slightly higher, yet over the same $C_n^2$ range the calculated THz BER changes by only ≈0.02 decades (≈$7.4\times10^{-7}$ to ≈$7.8\times10^{-7}$), two orders of magnitude less than the IR change. At this path length, the THz BER is therefore governed mainly by the receiver SNR, and the comparison constrains its level rather than the turbulence model.

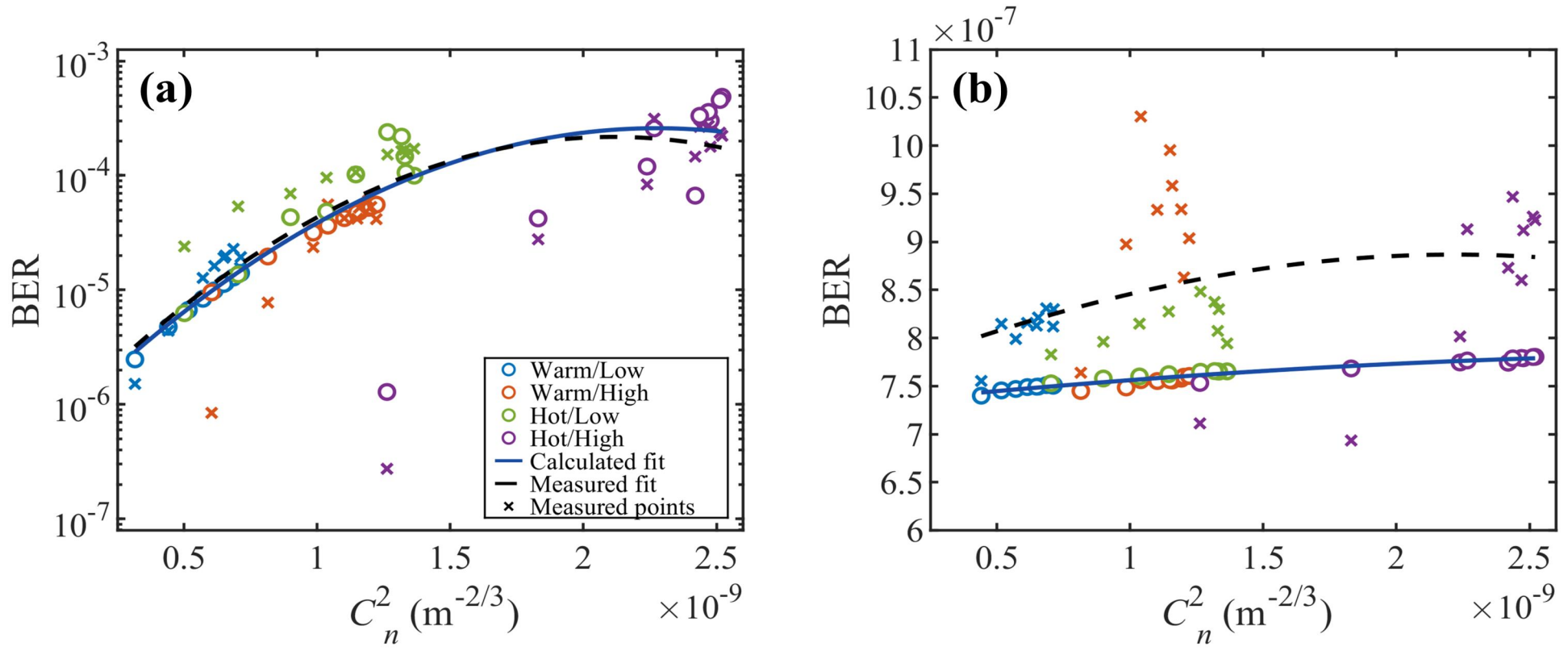

Fig. 4. Calculated and measured average BER versus $C_n^2$ for (a) the IR and (b) the THz link. Open circles and crosses denote calculated and measured one-second values, respectively; solid and dashed curves denote the corresponding fits in the $\log_{10}$(BER) domain.

## 5. Conclusion

In this work, co-propagating 625 GHz THz and 1550 nm IR communication links were measured in laboratory-controlled turbulence, while an MZI and thermistors independently retrieved the effective spectral exponent and the refractive-index structure parameter along the shared path. The measured turbulence parameters were used as inputs to Kolmogorov (K) and generalized non-Kolmogorov (NK) propagation models, and the predicted attenuation and BER were compared with the synchronized link measurements. Both the inlet-air temperature and the airflow speed increased the turbulence strength, but the temperature primarily controlled the transition from Kolmogorov-like to non-Kolmogorov turbulence.

For the IR link, the K model better reproduced the measured attenuation under the warm, Kolmogorov-like conditions, whereas the NK model performed better under the hot, non-Kolmogorov conditions at the nominal inner scale, even when the turbulence strengths were comparable. In contrast, the THz attenuation remained approximately 0.005-0.030 dB and was insensitive to the spectral type. With the type-matched propagation models and lognormal irradiance statistics, the predicted and measured BER trends agreed well, with the THz BER governed mainly by the receiver SNR. Under the tested short-path, weak-fluctuation conditions, these results indicate that IR link prediction requires the turbulence spectral shape and inner scale in addition to the turbulence strength, whereas THz fading can be budgeted without spectral characterization, favoring the THz link as a backup or control channel in hybrid THz/IR systems. Future work will investigate longer outdoor paths and stronger turbulence, with direct inner-scale measurements to better constrain the models.